\documentclass[aps,preprintnumbers, amsmath, amssymb]{revtex4}
\usepackage{braket} 
\usepackage{float}
\usepackage{graphics,epsfig}
\usepackage{graphicx}
\usepackage{dcolumn}
\usepackage{bm}
\usepackage{subfigure}
\usepackage{mathtools}
\begin{document}
\baselineskip=0.8 cm

\title{{\bf Effect of environment-induced interatomic interaction on entanglement generation for uniformly accelerated atoms with a boundary}}
\author{Chenhao Ma, Zixu Zhao\footnote{Corresponding author. zhao$_{-}$zixu@yeah.net}}
\affiliation{School of Science, Xi'an University of Posts and Telecommunications, Xi'an 710121, China}
\vspace*{0.2cm}
\begin{abstract}
\baselineskip=0.6 cm
\begin{center}
		{\bf Abstract}
\end{center}	
	
Considering environment-induced interatomic interaction, we study the entanglement dynamics of two uniformly accelerated atoms that interact with fluctuating massless scalar fields in the Minkowski vacuum in the presence of a reflecting boundary. The two atoms are initially prepared in a state such that one is in the ground state and the other is in the excited state, which is separable. When the acceleration is small, the rate of entanglement generation at the initial time and the maximum of concurrence generated during evolution oscillate with the distance between the atoms and the boundary before reaching a stable value, and may decrease non-monotonically with the acceleration, which means the anti-Unruh phenomenon can exist for some situations even when environmental considerations are taken into account. The results show that there exists the competition of the vacuum fluctuations caused by the boundary and the acceleration. In addition, the time evolution of concurrence will not be affected by the environment-induced interatomic interaction under certain conditions. For a larger acceleration, when the environment-induced interatomic interaction is considered, the concurrence may disappear later compared with the result when the environment-induced interatomic interaction is neglected.

\end{abstract}
\maketitle
\newpage
\vspace*{0.2cm}
\section{Introduction}

In 1935, Einstein, Podolsky and Rosen (EPR) attempted to challenge the completeness of quantum mechanics by a thought experiment \cite{Einstein1935}, which is the well-known EPR paradox. Bohr responded promptly to the paper \cite{Bohr}. Schr\"{o}dinger used ``Verschr\"{a}nkung" \cite{Schrodinger193501} and ``entanglement" \cite{Schrodinger193502} to discuss the phenomenon, which was considered ``spukhafte Fernwirkung" (``spooky action at a distance"). In order to solve the issue, Bell proposed the inequality \cite{Bell1964}, and the Bell's inequality was generalized \cite{Clauser1969}. The inequalities' violation means the existence of the strong correlations which cannot be accounted for by local realistic theory. The experiments of test Bell's theorem have been performed \cite{Freedman&Clauser:1972, Aspect198201,Aspect198202}. Thereafter, quantum entanglement has been studied widely. The utility of entangled states is by no means limited to testing the completeness of fundamental theories. Ekert presented a method in which the security of the so-called key distribution process in cryptography depends on the completeness of quantum mechanics \cite{Ekert1991}. It is recognized that quantum entanglement is a crucial resource in quantum physics. Jozsa \emph{et al.} demonstrated that two spatially separated parties can utilize shared prior quantum entanglement, and classical communications, to establish a synchronized pair of atomic clocks \cite{Jozsa2000}.

It is of great interest to consider the entanglement dynamics of non-inertial atoms. The Unruh effect \cite{Rindler,Fulling,Hawking1,Hawking2,Davies1,DeWitt,Unruh} states that, uniformly accelerated detector perceives the Minkowski vacuum as a thermal bath at a temperature proportional to its proper acceleration.  Benatti and Floreanini discussed the entanglement generation in uniformly accelerating atoms \cite{Benatti}. With the presence of a boundary, the accelerated atoms exhibit distinct features from static ones in a thermal bath at the corresponding Unruh temperature in terms of the entanglement creation at the neighborhood of the initial time \cite{ZhangYu2007}. The accelerated atoms can get entangled in certain circumstances while the inertial ones in the Minkowski vacuum cannot \cite{HuYu2015}, and the maximal concurrence generated during evolution for an initially separable two-atom system may increase with acceleration, which can be considered as an anti-Unruh effect in terms of the entanglement generated. On the other hand, entanglement harvesting has been investigated widely \cite{Salton2015,Henderson2018,Liu,Barman2021,Chowdhury2022,Barman2022,Barman2023,Barman&Majhi2023}, such as the effects of boundary and thermal bath.

Every quantum system is an open quantum system because it inevitably interacts with the environment at least with the vacuum fluctuations. The environment-induced interatomic interaction was generally neglected for dealing with the entanglement dynamics for accelerated atoms. Recently, Chen, Hu and Yu studied the influence of the environment-induced interatomic interaction on the entanglement dynamics of two uniformly accelerated atoms coupled with fluctuating massless scalar fields in the Minkowski vacuum, and showed that the anti-Unruh phenomenon in terms of the entanglement generation is deprived of by the consideration of the environment-induced interatomic interaction \cite{ChenHuYu}. As an inevitable result of the uncertainty principle, vacuum fluctuates and the modifications may have rich structures. The observed features of the detector model are closely related to the vacuum fluctuations of the field. The distortion of vacuum fluctuations can be caused by the acceleration of the detector, such as the Unruh effect. The vacuum fluctuations can also be modified by boundaries and the resulting distortions can produce novel observable effects, such as the Casimir effect. Adding a reflecting boundary, the vacuum fluctuations will be modified and we want to know how the entanglement will be affected.

The organization of the work is as follows. In Sec. II, we review basic formulas for two uniformly accelerated two-level atoms interacting with vacuum scalar fields. In Sec. III, we will consider the effect of environment-induced interatomic interaction on entanglement generation for uniformly accelerated atoms with a reflecting boundary. We conclude in the last section with our main results. We employ the natural units $\hbar = c = 1$ for convenience.

\section{The Basic Formulas}
		
In this section, we consider two uniformly accelerated two-level atoms interacting with real massless scalar fields. The Hamiltonian of the two atoms $H_{S}$ has the following form
\begin{equation}\label{hs}
	H_{S}=\frac{\omega}{2}\sigma_{3}^{(1)}+\frac{\omega}{2}\sigma_{3}^{(2)}.
\end{equation}
Here $\sigma^{(1)}_{i}=\sigma_{i}\otimes\sigma_{0}$ and $\sigma^{(2)}_{i}=\sigma_{0}\otimes\sigma_{i}$ are operators of atom $1$ and $2$ respectively. $\sigma_{0}$ is the $2\times2$ unit matrix and $\sigma_{i}~(i=1,2,3)$ the Pauli matrices. The transition frequencies of the two atoms are the same and denoted by $\omega$.
The interaction Hamiltonian $H_{I}$ between the two atoms and the vacuum scalar fields can be expressed similarly to the atom-light interaction as \cite{Audretsch1994}
\begin{equation}\label{hi}
	H_{I}=\mu[\sigma^{(1)}_{2}\Phi(t,x_{1})+\sigma^{(2)}_{2}\Phi(t,x_{2})],
\end{equation}
where $\mu$ is the small coupling constant. $\Phi(t,x_{\alpha}) (\alpha = 1, 2 )$ is the scalar field operator. We assume that the atoms are uncorrelated with the environment at the beginning. The initial state takes the form $\rho_{tot}(0)=\rho(0)\otimes|0\rangle\langle0|$, where $|0\rangle$ is the Minkowski vacuum state of the scalar fields and $\rho(0)$ is the initial state of the two-atom system. The master equation describing the dissipative dynamics of the two-atom subsystem in the weak-coupling limit can be expressed in the Gorini-Kossakowski-Lindblad-Sudarshan form as \cite{Kossakowski,Lindblad,Breure}
\begin{equation}\label{master1}
	\frac{\partial\rho(\tau)}{\partial\tau}=-i[H_{\rm eff},\rho(\tau)]+
	\mathcal{L}[\rho(\tau)],
\end{equation}
where the effective Hamiltonian is
\begin{equation}\label{master2}
	H_{\rm eff}=H_{S}-\frac{i}{2}\sum^{2}_{\alpha,\beta=1}\sum^{3}_{i,j=1}
	H_{ij}^{(\alpha\beta)}\sigma_{i}^{(\alpha)}\sigma_{j}^{(\beta)},
\end{equation}
and the dissipator takes
\begin{equation}\label{master3}
	\mathcal{L}[\rho(\tau)]=\frac{1}{2}\sum^{2}_{\alpha,\beta=1}\sum^{3}_{i,j=1}
C_{ij}^{(\alpha\beta)}[2\sigma_{j}^{(\beta)}\rho\sigma_{i}^{(\alpha)}-
\sigma_{i}^{(\alpha)}\rho\sigma_{j}^{(\beta)}-\rho\sigma_{i}^{(\alpha)}
\sigma_{j}^{(\beta)}].
\end{equation}
The environment leads to decoherence and dissipation depicted by the dissipator $\mathcal{L}[\rho(\tau)]$.  The coefficients of the matrix $C_{ij}^{(\alpha\beta)}$ and $H_{ij}^{(\alpha\beta)}$ are determined by the Fourier transform and Hilbert transform, $\mathcal{G}^{(\alpha\beta)}(\lambda)$ and $\mathcal{K}^{(\alpha\beta)}(\lambda)$, of the scalar field correlation function $\langle\Phi(\tau,x_{\alpha})\Phi(\tau',x_{\beta})\rangle$. The Fourier transform is
\begin{equation}\label{gk}
	\mathcal{G}^{(\alpha\beta)}(\lambda)=\int^{\infty}_{-\infty}d\Delta\tau
	e^{i\lambda\Delta\tau}\langle\Phi(\tau,x_{\alpha})\Phi(\tau',x_{\beta})\rangle,
\end{equation}
and the Hilbert transform is
\begin{align}\label{gwhl}
	\mathcal{K}^{(\alpha\beta)}(\lambda)=\frac{P}{\pi i}\int^{\infty}_{-\infty}d\omega
	\frac{\mathcal{G}^{(\alpha\beta)}(\omega)}{\omega-\lambda},
\end{align}
where $P$ is the principal value. The coefficients $C_{ij}^{(\alpha\beta)}$ in the dissipator (\ref{master3}) can be expressed as
\begin{equation}\label{cij}
	C_{ij}^{(\alpha\beta)}=A^{(\alpha\beta)}\delta_{ij}-iB^{(\alpha\beta)}
	\epsilon_{ijk}\delta_{3k}-A^{(\alpha\beta)}\delta_{3i}\delta_{3j},
\end{equation}
where
\begin{align}\label{AB}
	A^{(\alpha\beta)}=\frac{\mu^{2}}{4}[\mathcal{G}^{(\alpha\beta)}(\omega)+\mathcal{G}^{(\alpha\beta)}(-\omega)],\nonumber\\
	B^{(\alpha\beta)}=\frac{\mu^{2}}{4}[\mathcal{G}^{(\alpha\beta)}(\omega)-\mathcal{G}^{(\alpha\beta)}(-\omega)].
\end{align}
Similarly, $H_{ij}^{(\alpha\beta)}$ can be obtained by replacing the Fourier transform $\mathcal{G}^{(\alpha\beta)}(\lambda)$ with the Hilbert transform $\mathcal{K}^{(\alpha\beta)}(\lambda)$.

The effective Hamiltonian ${ H_{\rm eff}}$ can be written as ${ H_{\rm eff}}=\tilde{H_{s}}+{ H^{(12)}_{\rm eff}}$. $\tilde{H_{s}}$ is the renormalization of the transition frequencies. It has the same form as Eq. (\ref{hs}), but redefines the energy level spacing
\begin{equation}\label{omeg}
	\tilde{\omega}=\omega-\frac{i\mu^{2}}{2}[\mathcal{K}^{(11)}(\omega)-
	\mathcal{K}^{(11)}(-\omega)].
\end{equation}
	${ H^{(12)}_{\rm eff}}$ denotes the environment-induced coupling between the two atoms, and it takes the form
\begin{equation}\label{h12}
	H^{(12)}_{\rm eff}=-\sum_{i,j=1}^{3}\Omega_{ij}^{(12)}(\sigma_{i}\otimes\sigma_{j}),
\end{equation}
    where
\begin{equation}\label{omegaij}
	\Omega_{ij}^{(12)}=\mathcal{D}^{(12)}(\delta_{ij}-\delta_{3i}\delta_{3j}).
\end{equation}
$\mathcal{D}^{12}$ is the coefficient of $\Omega_{ij}^{(12)}$, and it takes the form $ \mathcal{D}^{(12)}=\frac{i\mu^{2}}{4}[\mathcal{K}^{(12)}(\omega)+\mathcal{K}^{(12)}(-\omega)]$. The master equation therefore can be rewritten in the form
\begin{align}\label{master}
	\nonumber
	\frac{\partial\rho(\tau)}{\partial\tau}=&-i\tilde{\omega}\sum_{\alpha=1}^{2}\,
	[\sigma_{3}^{(\alpha)},\rho(\tau)]+i
	\sum_{i,j=1}^{3}\Omega_{ij}^{(12)}[\sigma_{i}\otimes\sigma_{j},\rho(\tau)]\nonumber\\
	&+\frac{1}{2}\sum^{2}_{\alpha,\beta=1}\sum^{3}_{i,j=1}
	C_{ij}^{(\alpha\beta)}[2\sigma_{j}^{(\beta)}\rho\sigma_{i}^{(\alpha)}-
	\sigma_{i}^{(\alpha)}\rho\sigma_{j}^{(\beta)}-\rho\sigma_{i}^{(\alpha)}
	\sigma_{j}^{(\beta)}].
\end{align}
It should be noted that the second term contains environment-induced interatomic interaction.	

\section{Entanglement dynamics for accelerated atoms with a boundary}

\begin{figure}[htbp]
	\begin{center}	
		\includegraphics[scale=0.7]{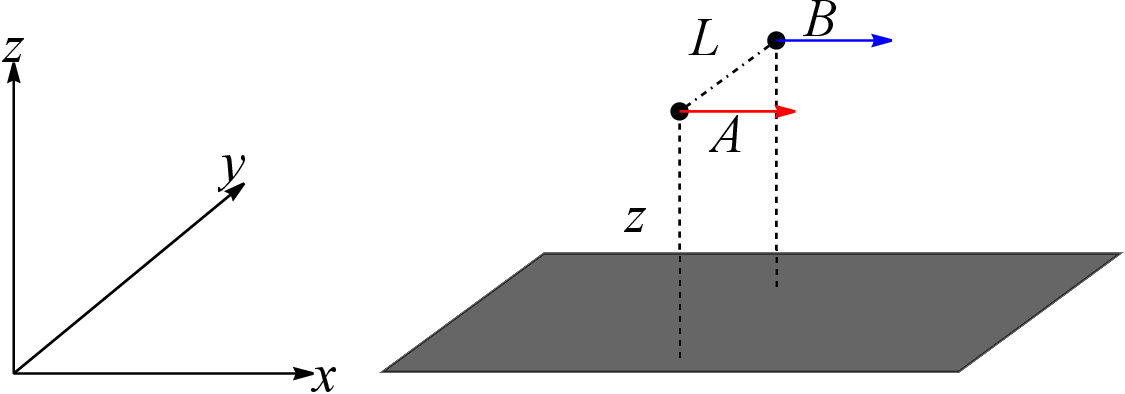}	
		\caption{\label{trajectory} The two atoms are separated with a distance $L$ and the reflecting boundary is located at $z=0$.}
	\end{center}
\end{figure}

In this section, we study the entanglement dynamics of a two-atom system in the presence of a reflecting boundary. We investigate how the entanglement dynamics is affected by the boundary and the environment-induced interatomic interaction. The two atoms are accelerated uniformly with the same acceleration along the $x$ axis, with a distance $z$ from the boundary, and are separated with a distance $L$ along the $y$ axis (Fig. \ref{trajectory}). The trajectories of the two uniformly accelerated atoms can be described as

\begin{align}\label{ab-tr}
	\nonumber t_{1}(\tau)=\frac{1}{a}\sinh(a\tau),\;\;x_{1}(\tau)=\frac{1}{a}\cosh(a\tau),
	\;\;y_{1}(\tau)=0,\;\;z_{1}(\tau)=z,\\
	t_{2}(\tau)=\frac{1}{a}\sinh(a\tau),\;\;x_{2}(\tau)=\frac{1}{a}\cosh(a\tau),
	\;\;y_{2}(\tau)=L,\;\;z_{2}(\tau)=z,
\end{align}
where $a$ is the proper acceleration. The Wightman function can be expressed as~\cite{Birrell1984}
\begin{align}\label{wigh-1}
	W\left(x, x'\right)=&-\frac{1}{4 \pi^{2}}\Big[\frac{1}{(t-t'-i
	\epsilon)^{2}-(x-x')^{2}-(y-y')^{2}
	-(z-z')^{2}}\nonumber\\
	&-\frac{1}{(t-t'-i \epsilon)^{2}-(x-x')^{2}-(y-y')^{2}
		-(z+z')^{2}}\Big]\;.
\end{align}
We can obtain Fourier transforms of the correlation functions
\begin{equation}
\begin{aligned}\label{g11g12}
\mathcal{G}^{(11)}(\omega)=&\mathcal{G}^{(22)}(\omega)=\frac{\omega}{4\pi}
 \left( \coth\left(\pi\omega/a \right)+1 \right)
	 \Big[1-f(\omega ,z)\Big] ,\\
	\mathcal{G}^{(12)}(\omega)=&\mathcal{G}^{(21)}(\omega)=\frac{\omega }{4 \pi }\left(\coth \left(\pi \omega/a\right)+1\right)\Big[f\left(\omega ,L/2\right)-f\left(\omega, \sqrt{L^2/4+ z^2}\right)\Big]\;,
\end{aligned}
\end{equation}
where
\begin{align}\label{01}	
 f(\omega ,z)=&\frac{\sin [\frac{2 \omega}{a} \sinh ^{-1}(a z)]}{2 \omega z \sqrt{a^2 z^2+1}}.
\end{align}		
Then $A^{(11)}=A^{(22)}=A_{1}$, $A^{(12)}=A^{(21)}=A_{2}$, $B^{(11)}=B^{(22)}=B_{1}$, $B^{(12)}=B^{(21)}=B_{2}$, $\mathcal{D}^{(12)}={D}$, we obtain
\begin{align}\label{cij19}
	&C_{ij}^{(11)}=C_{ij}^{(22)}=A_{1}\delta_{ij}-i B_{1}\epsilon_{ijk}\delta_{3k}-A_{1}\delta_{3i}\delta_{3j},\\
	&C_{ij}^{(12)}=C_{ij}^{(21)}=A_{2}\delta_{ij}-i B_{2}\epsilon_{ijk}\delta_{3k}-A_{2}\delta_{3i}\delta_{3j},\\
	&\Omega^{(12)}_{ij}=D\delta_{ij}-D\delta_{3i}\delta_{3j},\label{cij21}
\end{align}
where
\begin{align}\label{A1A2B1B2D}
	\nonumber A_{1}=&\frac{\Gamma_{0}}{4} \coth \left({\pi \omega }/{a}\right) \Big[ 1-f(\omega ,z) \Big],\\
	\nonumber A_{2}=&\frac{\Gamma_{0}}{4} \coth \left({\pi \omega }/{a}\right) \Big[f\left(\omega ,L/2\right)-f\left(\omega , \sqrt{L^2/4+z^2}\right) \Big],\\
	\nonumber B_{1}=&\frac{\Gamma_{0}}{4} \Big[1-f(\omega ,z)\Big],\\
	\nonumber B_{2}=&\frac{\Gamma_{0}}{4} \Big[f\left(\omega ,L/2\right)-f\left(\omega , \sqrt{L^2/4+ z^2}\right)\Big],\\
	 {D}=&\frac{\Gamma_{0}}{4}\Big[h\left(\omega ,L/2\right)-h\left(\omega , \sqrt{L^2/4+ z^2}\right)\Big],
\end{align}
with
\begin{align}\label{02}
 h(\omega ,z)=\frac{\cos [\frac{2 \omega}{a} \sinh ^{-1}(a z)]}{2 \omega z \sqrt{a^2 z^2+1}},
\end{align}
and $\Gamma_{0}=\frac{\mu^{2}\omega}{2\pi}$ denotes the spontaneous emission rate for an inertial atom in the Minkowski vacuum. We work in the coupled basis $\{|G\rangle=|00\rangle, |A\rangle=\frac{1}{\sqrt{2}}(|10\rangle-|01\rangle), |S\rangle=\frac{1}{\sqrt{2}}(|10\rangle+|01\rangle), |E\rangle=|11\rangle\}$, then time evolution equations of the density matrix elements can be expressed as \cite{Ficek2002}
\begin{align}\label{pf2}
	\nonumber \rho_{GG}'=&-4(A_{1}-B_{1})\rho_{GG}+2(A_{1}+B_{1}-A_{2}-B_{2})\rho_{AA}
	+2(A_{1}+B_{1}+A_{2}+B_{2})\rho_{SS},\\
	\nonumber \rho_{EE}'=&-4(A_{1}+B_{1})\rho_{EE}+2(A_{1}-B_{1}-A_{2}+B_{2})\rho_{AA}
	+2(A_{1}-B_{1}+A_{2}-B_{2})\rho_{SS},\\
	\nonumber \rho_{AA}'=&-4(A_{1}-A_{2})\rho_{AA}+2(A_{1}-B_{1}-A_{2}+B_{2})\rho_{GG}
	+2(A_{1}+B_{1}-A_{2}-B_{2})\rho_{EE},\\
	\nonumber \rho_{SS}'=&-4(A_{1}+A_{2})\rho_{SS}+2(A_{1}-B_{1}+A_{2}-B_{2})\rho_{GG}
	+2(A_{1}+B_{1}+A_{2}+B_{2})\rho_{EE},\\
	\nonumber \rho_{AS}'=&-4(A_{1}+i D)\rho_{AS},\;\;\;\;\;\;\;\;\;\;\;\;\;\;\;\;\;\;\;\;\;
	\;\;\;\;\;\;\;\;\;\;\;\;\;\;\rho_{SA}'=-4(A_{1}-i D)\rho_{SA},\\
	\rho_{GE}'=&-4A_{1}\rho_{GE},\;\;\;\;\;\;\;\;\;\;\;\;\;\;\;\;\;\;\;\;\;
	\;\;\;\;\;\;\;\;\;\;\;\;\;\;\;\;\;\;\;\;\;\;\;\;\;\rho_{EG}'=-4A_{1}\rho_{EG},
\end{align}
where $\rho_{IJ}=\langle I|\rho|J\rangle,I,J\in\{G,E,A,S\}$ and $\rho_{IJ}'=\frac{\partial\rho_{IJ}(\tau)}{\partial\tau}$. We assume that the initial density matrix is of the X form, then the X structure will be retained during evolution. The concurrence \cite{Wootters1998} as a measurement of quantum entanglement of a two-atom system in the X state takes the form \cite{Ficek2004}

\begin{align}\label{pfc}
	C[\rho(\tau)]=\max\{0,K_{1}(\tau),K_{2}(\tau)\},
\end{align}
	where
\begin{align}\label{pfk}
	\nonumber K_{1}(\tau)=&\sqrt{[\rho_{AA}(\tau)-\rho_{SS}(\tau)]^{2}-[\rho_{AS}(\tau)-\rho_{SA}(\tau)]^{2}}-
	2\sqrt{\rho_{GG}(\tau)\rho_{EE}(\tau)},\\
	K_{2}(\tau)=&2|\rho_{GE}(\tau)|-\sqrt{[\rho_{AA}(\tau)+\rho_{SS}(\tau)]^{2}-
		[\rho_{AS}(\tau)+\rho_{SA}(\tau)]^{2}}.
\end{align}

\subsection{The rate of entanglement generation}

Considering the environment-induced interatomic interaction, we study the entanglement generation between two initially separable uniformly accelerated atoms in the presence of a reflecting boundary. From Eq. (\ref{pf2}), when we study the effect of environment-induced interatomic interaction on entanglement dynamics, the initial state cannot be chosen such that $\rho_{AS}(0)=\rho_{SA}(0) = 0$. In the following discussion, we assume that the initial state of the two-atom system is $|10\rangle$ [$\rho_{AS}(0)=\rho_{SA}(0)=\frac{1}{2}$].

The atoms are initially non-entangled and we study the entanglement generation near the initial time by means of the evolution equations (\ref{pf2}) and the expression of concurrence \eqref{pfc}.
When the initial state is chosen as $|10\rangle$, it is clear that $\rho_{GE}(\tau)\equiv0$, so $K_{2}(\tau)<0$, and the concurrence $C[\rho(\tau)]= \max\{0,K_{1}(\tau)\}$. Because $K_{1}(0)=0$, entanglement can be generated near the initial time $\tau=0$ when $K_{1}'(0)>0$.
We can obtain

\begin{align}\label{k11}
{K}_{1}'(0)=4\sqrt{A_{2}^{2}+D^{2}}-4\sqrt{A_{1}^{2}-B_{1}^{2}}.
\end{align}
Here $D$ contains the environment-induced interatomic interaction. From Eq.~(\ref{k11}), compared with that when the environment-induced interaction is neglected $(D=0)$, $C'(0)$ is always larger when the environment induced interaction is taken into account $(D\neq0)$. The condition for entanglement generation near the initial time is
\begin{align}\label{pf8}
A_{2}^{2}+D^{2}>A_{1}^{2}-B_{1}^{2}.
\end{align}
Entanglement is more likely to be generated when the environment-induced interaction is considered $(D\neq0)$ compared with the result when the environment-induced interaction is neglected $(D=0)$.

\begin{figure}[]
	\begin{center}	
		\includegraphics[width=0.35\textwidth]{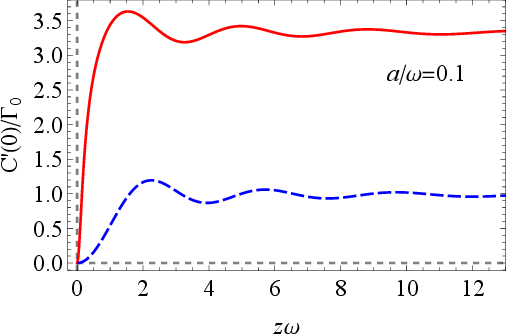}\hspace{0.02\textwidth}
		\includegraphics[width=0.35\textwidth]{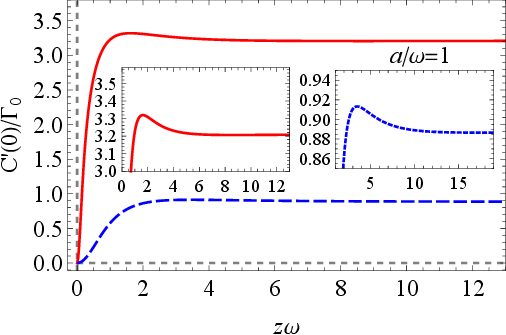}\hspace{0.02\textwidth}	
		\caption{\label{dcz} The rate of entanglement generation at the initial time is plotted as a function of $z\omega$ with $L\omega=0.3$ and $a/\omega=\{0.1, 1\}$. The solid(red) and dashed(blue) lines represent uniformly accelerated atoms with and without the environment-induced interatomic interaction, respectively.}
	\end{center}
\end{figure}

In Fig. \ref{dcz}, we describe the rate of entanglement generation at the initial time $C'(0)$ as a function of the distance between atoms and the boundary. For a small acceleration such as $a/\omega=0.1$, as $z\omega$ increases, $C'(0)$ initially increases, then oscillates, and finally reaches a stable value. The oscillation is caused by the modification of the vacuum fluctuations due to the presence of the reflecting boundary. For a larger $a/\omega$ such as $a/\omega=1$, $C'(0)$ first increases, then decreases to a stable value. The maximum value will occur near the boundary. The oscillation disappears for a larger acceleration. This is the result of the competition between the boundary effect and the acceleration effect. $C'(0)$ is always larger when the environment-induced interatomic interaction is non-zero compared with the result when the environment-induced interatomic interaction is neglected.
	
\begin{figure}[htbp]
	\begin{center}		
		\includegraphics[scale=0.55]{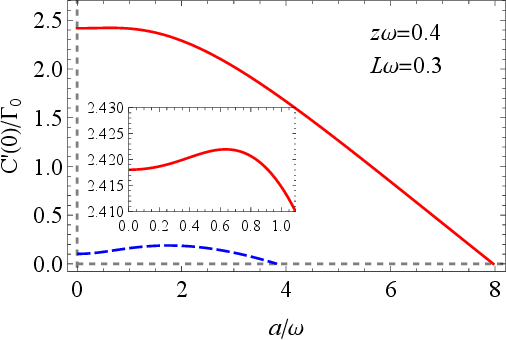}\vspace{0.01\textwidth}\hspace{0.02\textwidth}
		\includegraphics[scale=0.55]{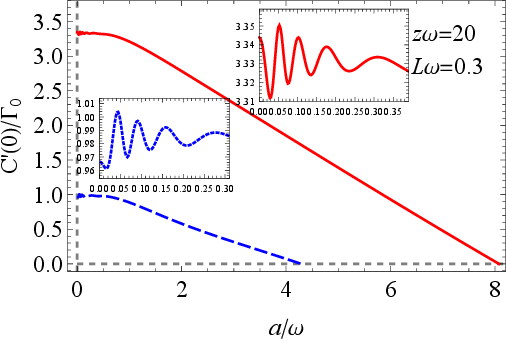}\vspace{0.01\textwidth}\hspace{0.02\textwidth}
		\includegraphics[scale=0.55]{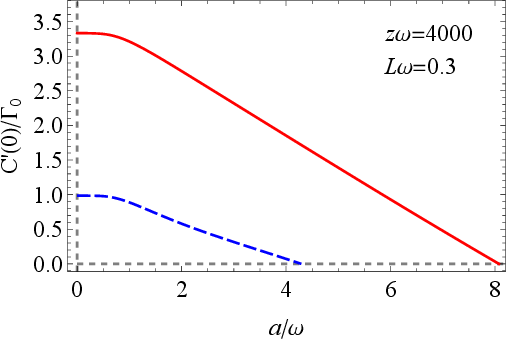}\vspace{0.01\textwidth}
		\includegraphics[scale=0.55]{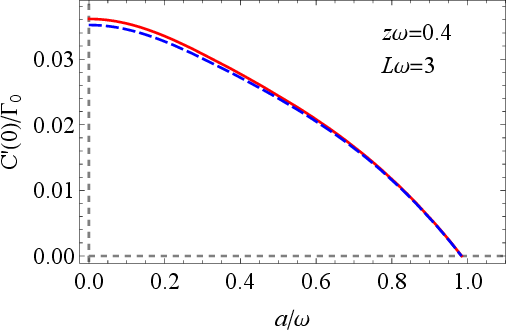}\vspace{0.01\textwidth}\hspace{0.02\textwidth}
		\includegraphics[scale=0.55]{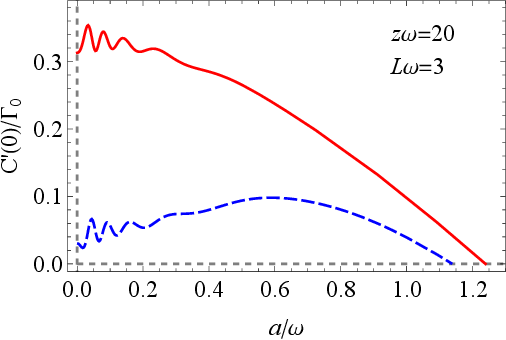}\vspace{0.01\textwidth}\hspace{0.02\textwidth}
		\includegraphics[scale=0.55]{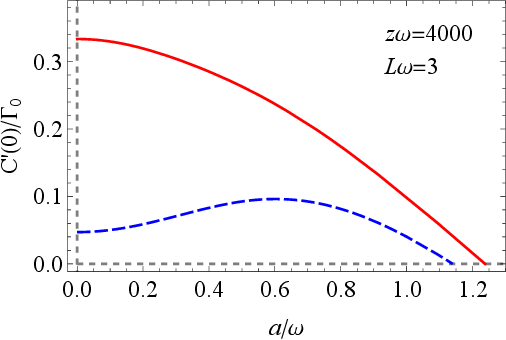}\vspace{0.01\textwidth}
		\includegraphics[scale=0.55]{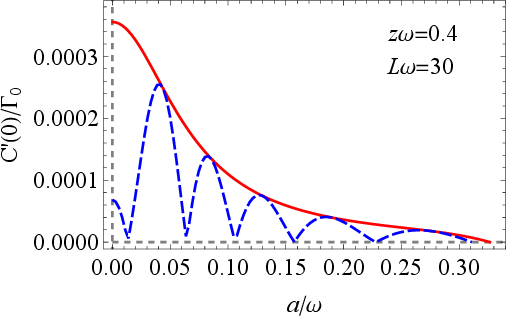}\vspace{0.01\textwidth}\hspace{0.02\textwidth}
		\includegraphics[scale=0.55]{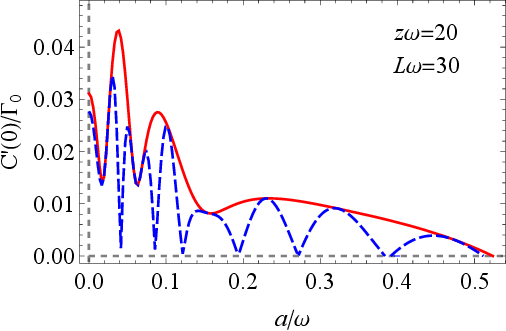}\hspace{0.02\textwidth}
		\includegraphics[scale=0.55]{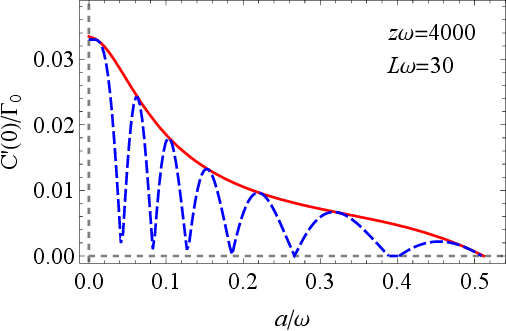}
        \caption{\label{dca} The rate of entanglement generation at the initial time is plotted as a function of $a/\omega$ with $z\omega=\{0.4, 20, 4000\}$ and $L\omega=\{0.3, 3, 30\}$. The solid(red) and dashed(blue) lines represent uniformly accelerated atoms with and without the environment-induced interatomic interaction, respectively.}
    \end{center}
\end{figure}

In Fig. \ref{dca}, we depict the rate of entanglement generation at the initial time $C'(0)$ as a function of acceleration. For a small $z\omega$, when the environment-induced interatomic interaction is considered, as acceleration increases, $C'(0)$ may first increase, and then decrease to zero, or may decrease monotonically with acceleration. For a large $L\omega$, $C'(0)$ displays oscillatory behavior when the environment-induced interatomic interaction is neglected. As $z\omega$ increases, $C'(0)$ shows an oscillatory behavior with acceleration, but the oscillatory behavior only exists at small acceleration for a not large $L\omega$. When the atoms are far from the boundary, $C'(0)$ decreases monotonically with acceleration when the environment-induced interatomic interaction is considered, which is similar to the free space \cite{ChenHuYu}.

\begin{figure}[htbp]
	\begin{center}	
		\includegraphics[scale=0.55]{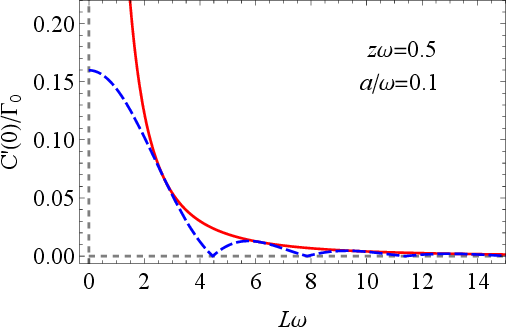}\vspace{0.01\textwidth}\hspace{0.02\textwidth}
		\includegraphics[scale=0.55]{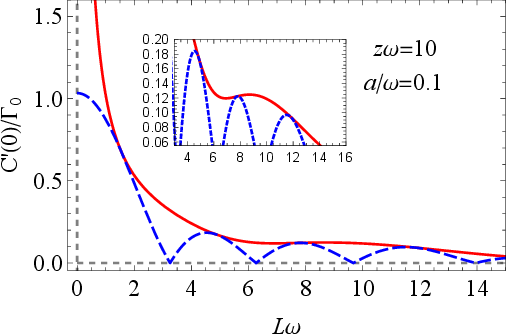}\vspace{0.01\textwidth}\hspace{0.02\textwidth}
		\includegraphics[scale=0.55]{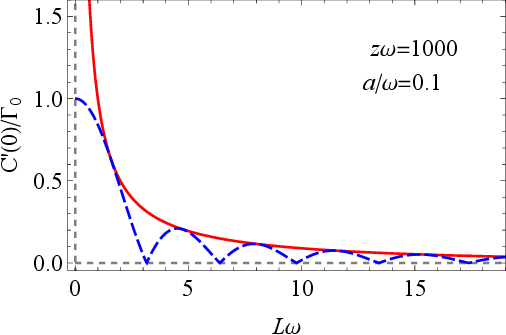}\vspace{0.01\textwidth}
        \includegraphics[scale=0.55]{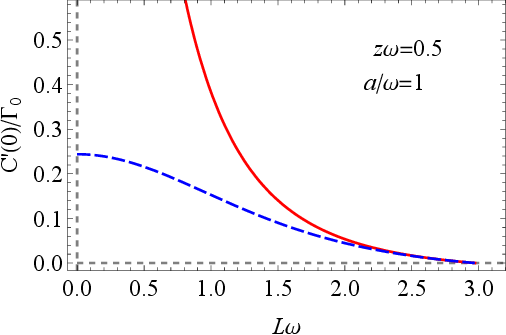}\vspace{0.01\textwidth}\hspace{0.02\textwidth}
		\includegraphics[scale=0.55]{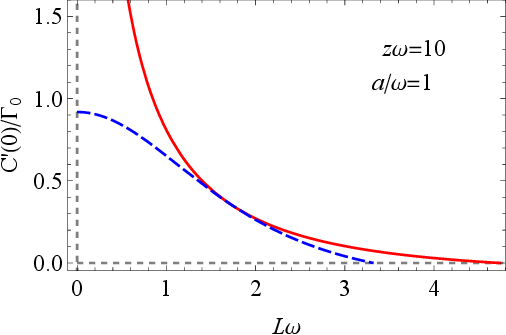}\vspace{0.01\textwidth}\hspace{0.02\textwidth}
		\includegraphics[scale=0.55]{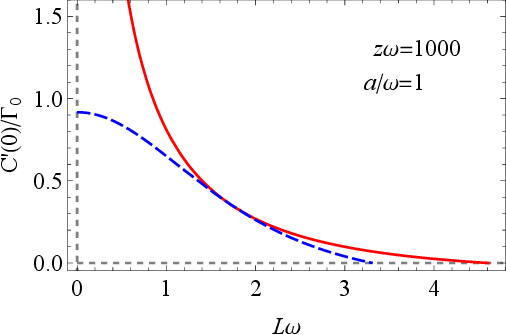}\vspace{0.01\textwidth}
		\caption{\label{dcL} The rate of entanglement generation at the initial time is plotted as a function of $L\omega$ with $a/\omega=\{0.1, 1\}$ and $z\omega=\{0.5, 10, 1000\}$. The solid(red) and dashed(blue) lines represent uniformly accelerated atoms with and without the environment-induced interatomic interaction, respectively.}
	\end{center}
\end{figure}

As shown in Fig.~\ref{dcL}, we plot the rate of entanglement generation at the initial time as a function of the interatomic separation $L\omega$. For a small $a/\omega$, $C'(0)$ exhibits oscillatory behavior with $L\omega$ when the environment-induced interatomic interaction is ignored. When the environment-induced interatomic interaction is considered, $C'(0)$ decreases monotonically with $L\omega$ for a small $z\omega$. As $z\omega$ increases, for a certain $z\omega$, the rate of entanglement generation at the initial time does not decrease monotonically with $L\omega$. For a large $z\omega$, the behavior is similar to the small $z\omega$ case. For a larger $a/\omega$, $C'(0)$ decreases monotonically with $L\omega$ when the environment-induced interatomic interaction is considered, which is similar to the free space case \cite{ChenHuYu}. Similar studies were done on black hole spacetimes in \cite{Barman202201,Barman2024}. The concurrence may vary periodically with the distance between the two null paths of the detectors \cite{Barman202201}.

\subsection{The evolution process of entanglement}

We study the time evolution of entanglement with a two-atom system initially prepared in the state $|10\rangle$. From Eq. (23), the expression for the density matrix element $\rho_{AS}(\tau)$ is,

\begin{align}\label{ras}
\rho_{AS}(\tau)=\rho_{SA}^*(\tau)=\frac{1}{2} e^{-4(A_{1}+iD)\tau}.
\end{align}

From Eqs. (\ref{pfc})-(\ref{pfk}), the concurrence is $C[\rho(\tau)]=\max\{0,K_{1}(\tau)\}$, where

\begin{align}\label{k1}
K_{1}(\tau)=\sqrt{[\rho_{AA}(\tau)-\rho_{SS}(\tau)]^{2}+\sin^{2}(4D\tau)e^{-8A_{1}\tau}}-2\sqrt{\rho_{GG}(\tau)\rho_{EE}(\tau)}.
\end{align}

\begin{figure}[htbp]
\begin{center}
    \includegraphics[scale=0.7]{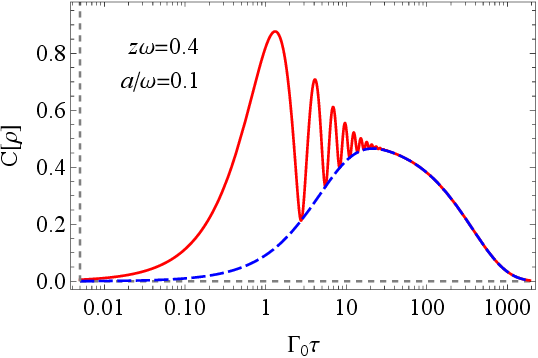}\vspace{0.01\textwidth}\hspace{0.02\textwidth}
	\includegraphics[scale=0.7]{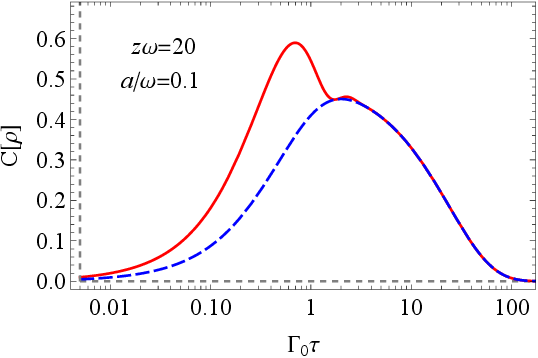}\vspace{0.01\textwidth}
	\includegraphics[scale=0.7]{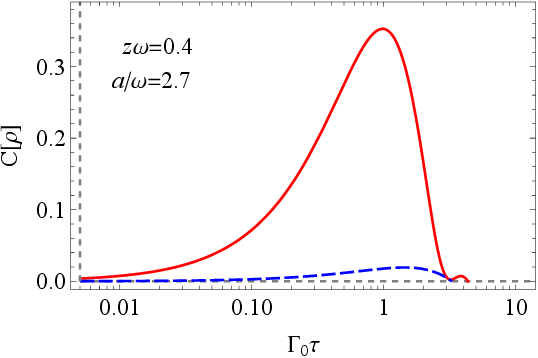}\hspace{0.02\textwidth}		
	\includegraphics[scale=0.7]{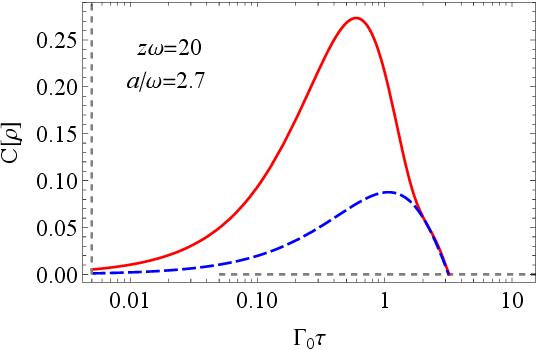}
\caption{\label{CvstL05} The time evolution of concurrence for $L\omega=0.5$ with $z\omega=\{0.4, 20\}$ and $a/\omega=\{0.1, 2.7\}$. The solid(red) and dashed(blue) lines represent uniformly accelerated atoms with and without the environment-induced interatomic interaction initially prepared in $|10\rangle$, respectively.}
\end{center}
\end{figure}

\begin{figure}[htbp]
\begin{center}
	\includegraphics[scale=0.55]{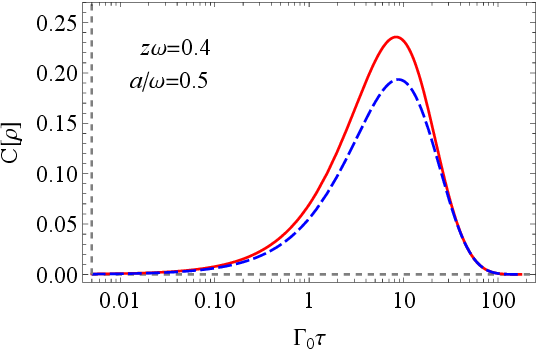}\vspace{0.01\textwidth}
	\includegraphics[scale=0.55]{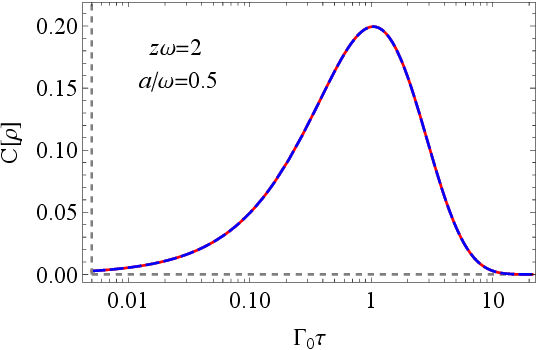}\vspace{0.01\textwidth}
	\includegraphics[scale=0.55]{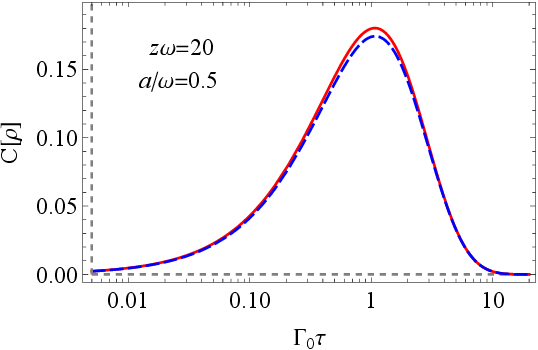}\vspace{0.01\textwidth}
	\includegraphics[scale=0.55]{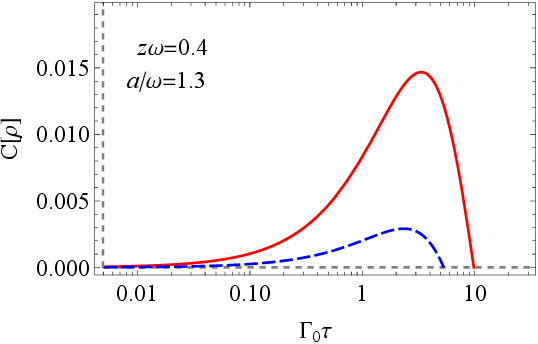}\vspace{0.01\textwidth}
	\includegraphics[scale=0.55]{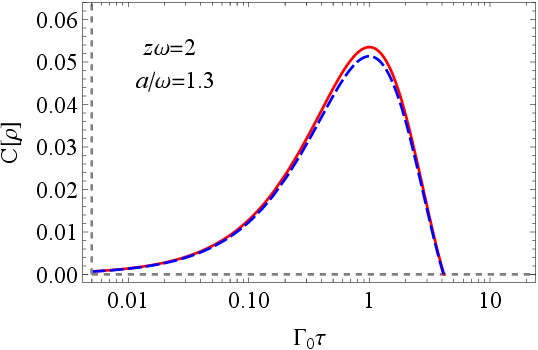}	
	\includegraphics[scale=0.55]{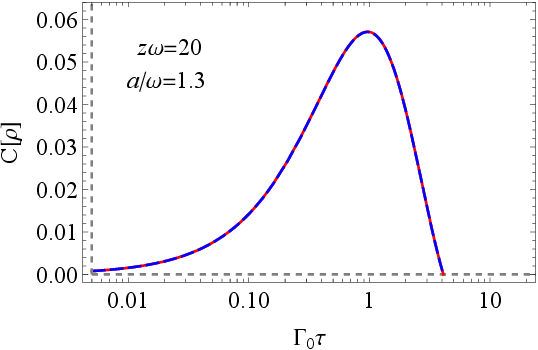}
	\caption{\label{CvstL1d9} The time evolution of concurrence for $L\omega=1.9$ with $z\omega=\{0.4, 2, 20\}$ and $a/\omega=\{0.5, 1.3\}$. The solid(red) and dashed(blue) lines represent uniformly accelerated atoms with and without the environment-induced interatomic interaction initially prepared in $|10\rangle$, respectively.}
\end{center}
\end{figure}

From Eq. (\ref{k1}), considering the environment-induced interatomic interaction, there is an additional term $\sin^{2}(4D\tau)\,e^{-8A_{1}\tau}$. Therefore, oscillation appears in the time evolution of concurrence, and the oscillation is damped during the evolution so that the asymptotic concurrence is not affected by the environment-induced interatomic interaction. For $4D\tau=n\pi \{n=0,1,2,3...\}$, the additional term is zero. Therefore, the environment-induced interatomic interaction will disappear for appropriate parameters.

In Figs.~\ref{CvstL05} and \ref{CvstL1d9}, we study the time evolution of concurrence. For a larger acceleration, when the environment-induced interatomic interaction is considered, $C[\rho]$ may disappear later compared with the result when the environment-induced interatomic interaction is neglected. For a small $L\omega$, as shown in Fig. \ref{CvstL05}, when the environment-induced interatomic interaction is considered, $C[\rho]$ initially increases, then oscillates, and finally asymptotically approaches the case without environment-induced interatomic interaction. The oscillatory behavior is more obvious for small $a/\omega$ and small $z\omega$. For a larger $L\omega$, the oscillation disappears as shown in Fig. \ref{CvstL1d9}. $C[\rho]$ first increases, and then decreases to zero. The two curves coincide exactly for some conditions, which means the concurrence will not be affected by the environment-induced interatomic interaction in some circumstances.

\subsection{The maximum of concurrence generated during evolution}

\begin{figure}[htbp]
\begin{center}
	\includegraphics[scale=0.7]{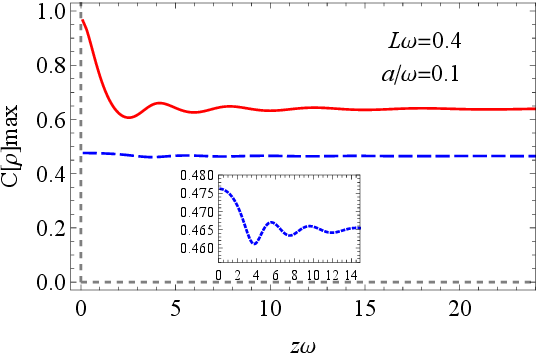}\vspace{0.01\textwidth}\hspace{0.02\textwidth}
	\includegraphics[scale=0.7]{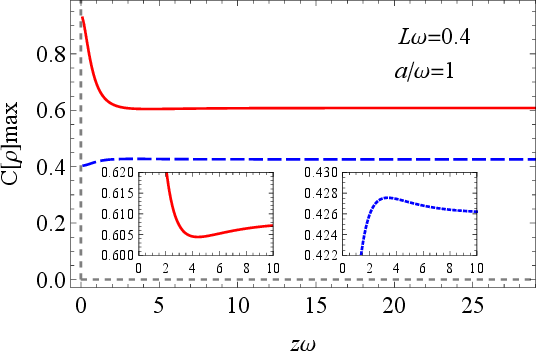}\vspace{0.01\textwidth}
	\caption{\label{CmaxvszLsmall}The maximum of concurrence during evolution $C[\rho]_{max}$ is plotted as a function of $z\omega$ with $L\omega=0.4$ and $a/\omega=\{0.1, 1\}$. The solid(red) lines and dashed(blue) lines represent uniformly accelerated atoms with and without the environment-induced interatomic interaction initially prepared in $|10\rangle$, respectively. }
\end{center}
\end{figure}

\begin{figure}[htbp]
\begin{center}
	\includegraphics[scale=0.7]{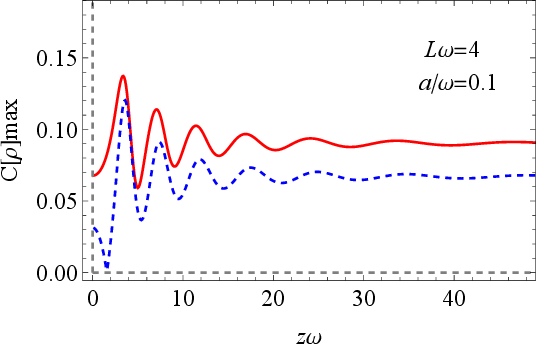}\vspace{0.01\textwidth}\hspace{0.02\textwidth}
	\includegraphics[scale=0.7]{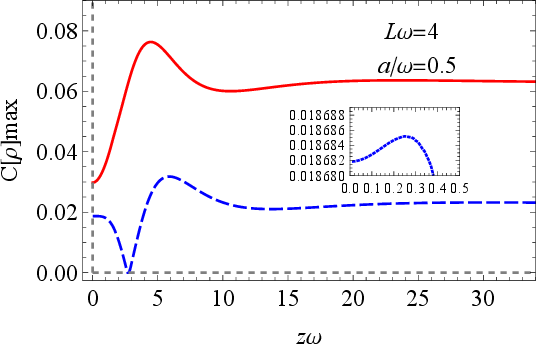}\vspace{0.01\textwidth}
	\includegraphics[scale=0.7]{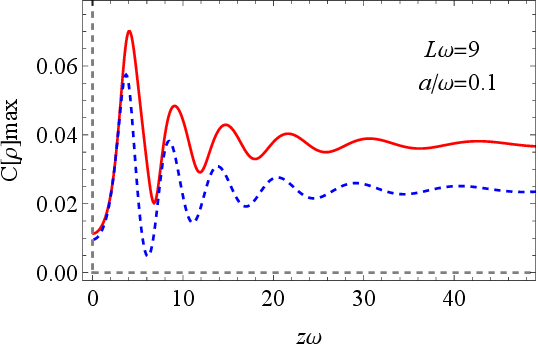}\hspace{0.02\textwidth}
	\includegraphics[scale=0.7]{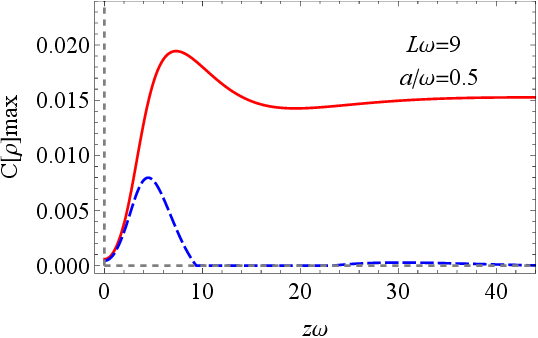}	
	\caption{\label{CmaxvszLbig}The maximum of concurrence during evolution $C[\rho]_{max}$ is plotted as a function of $z\omega$ with $L\omega=\{4, 9\}$ and $a/\omega=\{0.1, 0.5\}$. The solid(red) lines and dashed(blue) lines represent uniformly accelerated atoms with and without the environment-induced interatomic interaction initially prepared in $|10\rangle$, respectively.}
\end{center}
\end{figure}

We now consider the maximum of concurrence generated during evolution. In Figs. \ref{CmaxvszLsmall} and \ref{CmaxvszLbig}, we plot the maximum of concurrence generated during evolution $C[\rho]_{max}$ as a function of the distance between atoms and the boundary.

\begin{figure}[htbp]
	\begin{center}
		\includegraphics[scale=0.55]{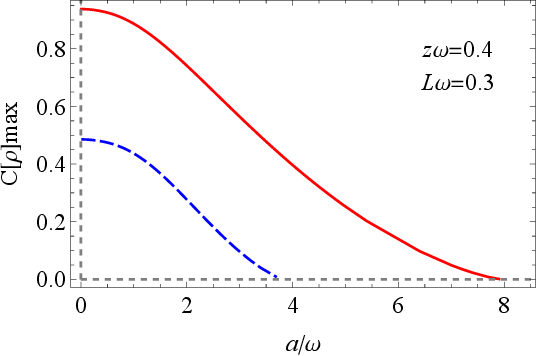}\vspace{0.01\textwidth}
		\includegraphics[scale=0.55]{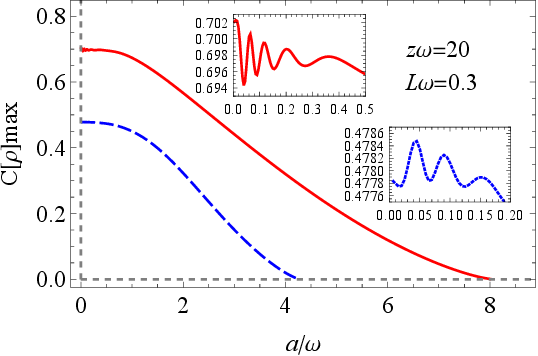}\vspace{0.01\textwidth}
		\includegraphics[scale=0.55]{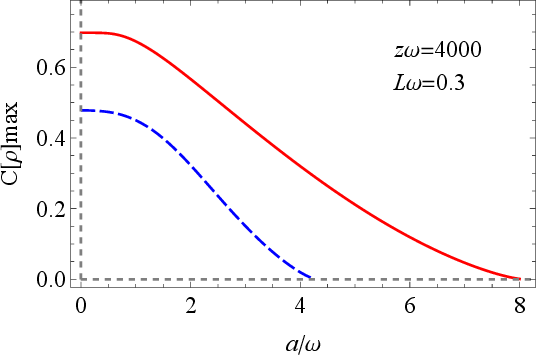}\vspace{0.01\textwidth}
		\includegraphics[scale=0.55]{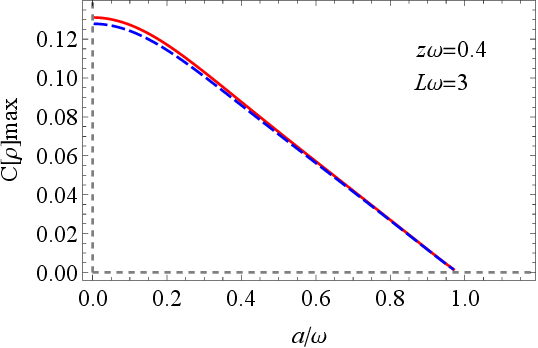}\vspace{0.01\textwidth}
		\includegraphics[scale=0.55]{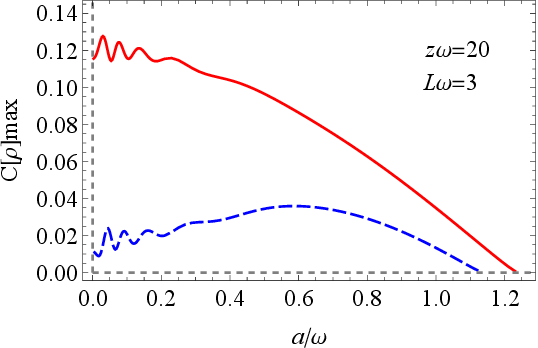}\vspace{0.01\textwidth}
		\includegraphics[scale=0.55]{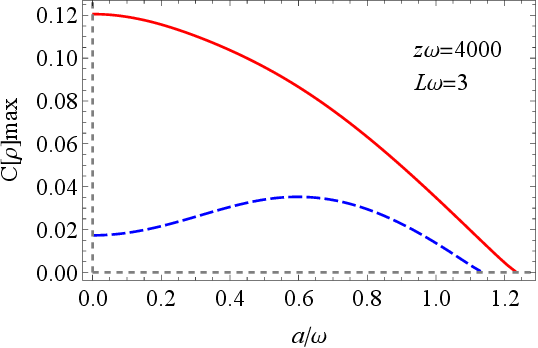}\vspace{0.01\textwidth}
		\includegraphics[scale=0.55]{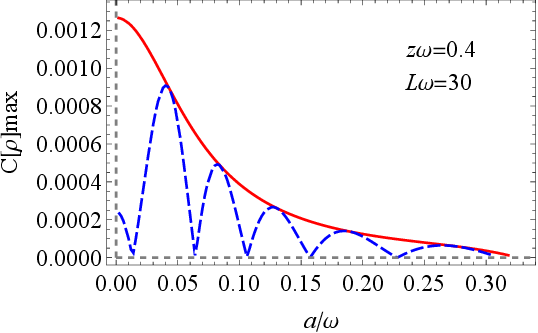}\vspace{0.01\textwidth}
		\includegraphics[scale=0.55]{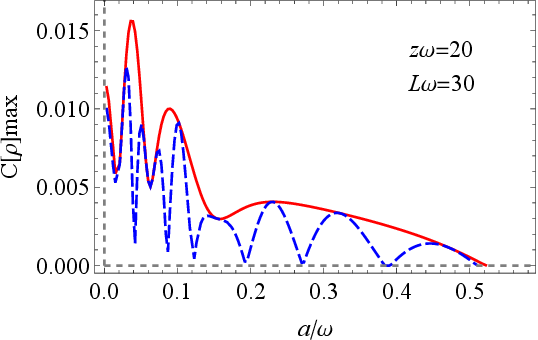}\vspace{0.01\textwidth}
		\includegraphics[scale=0.55]{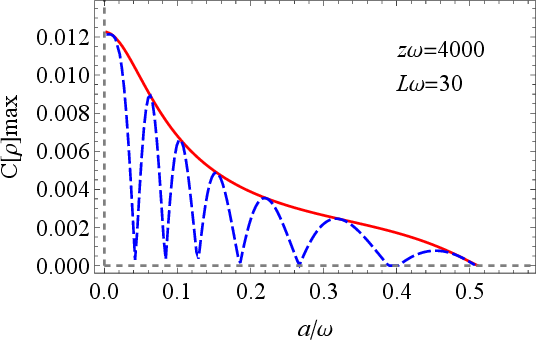}\vspace{0.01\textwidth}
		\caption{\label{Cmaxvsa}The maximum of concurrence during evolution $C[\rho]_{max}$ is plotted as a function of $a/\omega$ with $L\omega=\{0.3, 3, 30\}$ and $z\omega=\{0.4, 20, 4000\}$. The solid(red) lines and dashed(blue) lines represent uniformly accelerated atoms with and without the environment-induced interatomic interaction initially prepared in $|10\rangle$, respectively.}
	\end{center}
\end{figure}

For a small $L\omega$ as shown in Fig. \ref{CmaxvszLsmall}, with a small $a/\omega$, $C[\rho]_{max}$ first decreases, then oscillates, and finally arrives at a stable value. This indicates that it is more favorable near the boundary for entanglement generation with a small $L\omega$. With a larger $a/\omega$, when the environment-induced interatomic interaction is considered, $C[\rho]_{max}$ initially decreases, then increases, and finally reaches a stable value. There exists a minimal value. Ignoring environment-induced interatomic interaction, we observe that $C[\rho]_{max}$ first increases from a certain value, then decreases, and finally goes to a stable value. When we take a larger $L\omega$, as shown in Fig. \ref{CmaxvszLbig}, considering the environment-induced interatomic interaction, with a small $a/\omega$, we find that $C[\rho]_{max}$  first increases, then oscillates, and eventually achieves a stable value. Without the environment-induced interatomic interaction, $C[\rho]_{max}$ may increase or decrease at first. For a larger $a/\omega$, when the environment-induced interatomic interaction is considered, $C[\rho]_{max}$ first increases, then decreases, and finally increases to a stable value. When the environment-induced interatomic interaction is neglected, $C[\rho]_{max}$ may increase or decrease at first.

\begin{figure}[htbp]
	\begin{center}
		\includegraphics[scale=0.55]{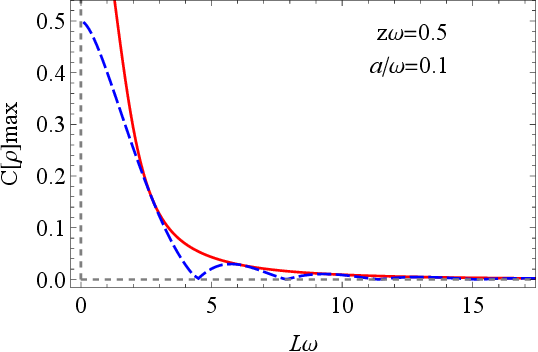}\vspace{0.01\textwidth}\hspace{0.02\textwidth}
		\includegraphics[scale=0.55]{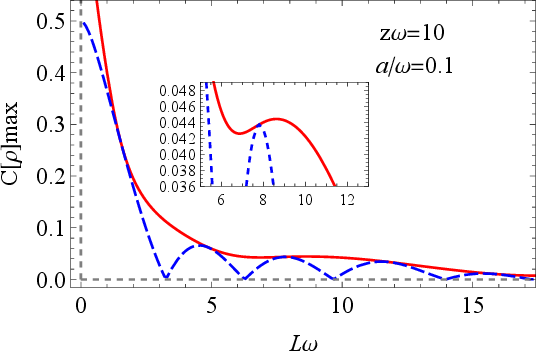}\vspace{0.01\textwidth}\hspace{0.02\textwidth}
		\includegraphics[scale=0.55]{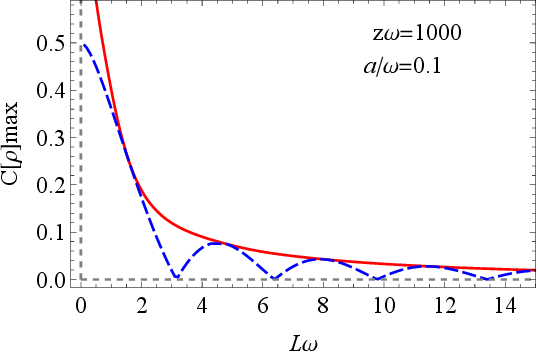}\vspace{0.01\textwidth}
		\includegraphics[scale=0.55]{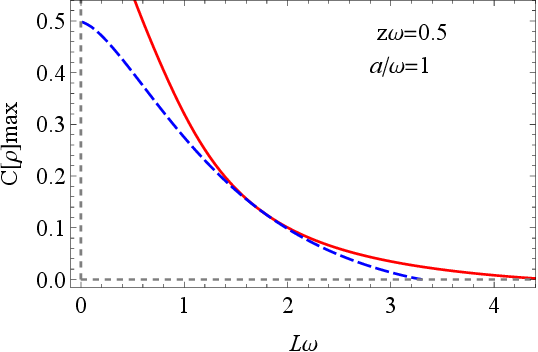}\vspace{0.01\textwidth}\hspace{0.02\textwidth}
		\includegraphics[scale=0.55]{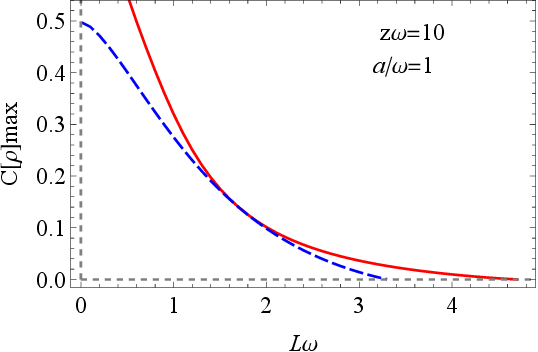}\vspace{0.01\textwidth}\hspace{0.02\textwidth}
		\includegraphics[scale=0.55]{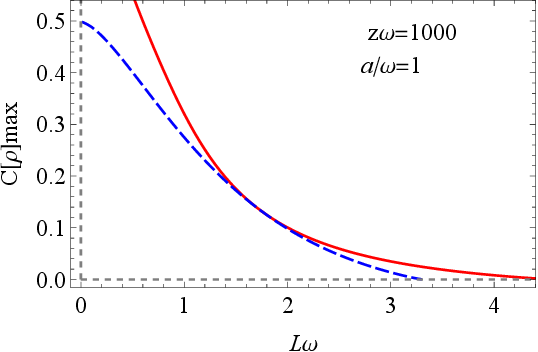}\vspace{0.01\textwidth}
		\caption{\label{CmaxvsL}The maximum of concurrence during evolution $C[\rho]_{max}$ is plotted as a function of $L\omega$ with $a/\omega=\{0.1, 1\}$ and $z\omega=\{0.5, 10, 1000\}$. The solid(red) lines and dashed(blue) lines represent uniformly accelerated atoms with and without the environment-induced interatomic interaction initially prepared in $|10\rangle$, respectively.}
	\end{center}
\end{figure}

We describe the maximum of concurrence during evolution as a function of acceleration in Fig. \ref{Cmaxvsa}. When the environment-induced interatomic interaction is considered, the maximum of concurrence during evolution decreases monotonically with acceleration for a small or large $z\omega$, but the maximum of the concurrence during evolution decreases non-monotonically with acceleration for a medium $z\omega$. Taking a large $L\omega$, the oscillation will be noticeable for the $C[\rho]_{max}$ when the environment-induced interatomic interaction is ignored. In Fig. \ref{CmaxvsL}, for a small acceleration, the maximum of concurrence generated during evolution may decrease non-monotonically to zero with $L\omega$ when the environment-induced interatomic interaction is considered. For a larger acceleration, the maximum concurrence generated during evolution decreases monotonically.

\section{Conclusion}

We have studied the entanglement dynamics of uniformly accelerated atoms coupled with fluctuating massless scalar fields with a boundary in the Minkowski vacuum when the environment-induced interatomic interaction is considered. The two atoms are initially prepared in the ground and excited states respectively, which is separable.

For a small acceleration, as the distance between the atoms and the boundary increases, the rate of entanglement generation at the initial time first increases, then oscillates, and finally arrives at a stable value. When we take a larger acceleration, it first increases, then decreases to a stable value. Regardless of whether the environment-induced interatomic interaction is considered, the rate of entanglement generation at the initial time oscillates with acceleration for some distance between the atoms and the boundary. When the environment-induced interatomic interaction is taken into account, the rate of entanglement generation may decrease non-monotonically with the interatomic separation for a certain distance between the atoms and the boundary.

For a fixed small interatomic separation, the time evolution of concurrence exhibits obvious oscillation for small acceleration and small distance between the atoms and the boundary. With a fixed larger interatomic separation, the time evolution of concurrence first increases, and then decreases to zero, and the concurrence will not be affected by the environment-induced interatomic interaction for some cases.

The maximum of concurrence generated during evolution is also studied. For a small acceleration, with and without the environment-induced interatomic interaction, the maximum of concurrence generated during evolution exhibits oscillatory behavior with the distance between the atoms and the boundary. For a not small distance between the atoms and the boundary, the maximum of concurrence generated during evolution may oscillate with the acceleration. For a small acceleration, the maximum of concurrence may decrease non-monotonically with the interatomic separation when the environment-induced interatomic interaction is considered. There are competing vacuum fluctuations caused by the boundary and the acceleration.

\begin{acknowledgments}

This work was supported by the Nature Science Foundation of Shaanxi Province, China under Grant No. 2023-JC-YB-016 and the National Natural Science Foundation of China under Grant No. 11705144.

\end{acknowledgments}


\end{document}